\documentclass[aps,prl,twocolumn,showpacs,superscriptaddress]{revtex4-1}

\usepackage{amsmath}    
\usepackage{graphicx}   
\usepackage{verbatim}   
\usepackage{color}      
\usepackage{hyperref}   
\usepackage{units}
\usepackage{braket}
\usepackage{siunitx}

\definecolor{grin}{RGB}{19,163,19}

\usepackage[dvipsnames,svgnames,x11names]{xcolor}

\usepackage[final]{changes}

\setauthormarkupposition{left}
\setremarkmarkup{(#2)}

\definechangesauthor[color=ForestGreen]{Fred}
\definechangesauthor[color=Orchid]{Valentin}
\definechangesauthor[color=Cerulean]{Vincent}
\definechangesauthor[color=Dandelion]{Laurent}
\definechangesauthor[color=Bittersweet]{Alain}

\newcommand{\tauS}{\tau_\mathrm{s}}
\newcommand{\tauSB}{\tau_\mathrm{s}^\mathrm{Born} }

\newcommand{\lS}{l_\mathrm{s}}
\newcommand{\lSB}{l_\mathrm{s}^\mathrm{Born}  }

\newcommand{\VR}{V_\mathrm{R}}

\newcommand{\ki}{\mathbf{k}_i}
\newcommand{\kp}{\mathbf{k}'}
\newcommand{\kdis}{\mathbf{k}_\mathrm{dis}}

\newcommand{\sperp}{\sigma}

\newcommand{\old}[1]{\textcolor{Blue}{\sout{#1}}}
\newcommand{\comm}[1]{\textcolor{Gray}{#1}}
\renewcommand{\old}[1]{}
\renewcommand{\comm}[1]{}

\begin{document}

\title{Elastic Scattering Time of Matter-Waves in Disordered Potentials}

\author{J\'{e}r\'{e}mie Richard} 
\affiliation{Laboratoire Charles Fabry, Institut d'Optique, CNRS, Universit\'{e} Paris-Saclay, 91127 Palaiseau cedex, France}

\author{Lih-King Lim} 
\affiliation{Zhejiang Institute of Modern Physics, Zhejiang University, Hangzhou 310027, P. R. China}
\affiliation{Laboratoire Charles Fabry, Institut d'Optique, CNRS, Universit\'{e} Paris-Saclay, 91127 Palaiseau cedex, France}

\author{Vincent Denechaud} 
\affiliation{Laboratoire Charles Fabry, Institut d'Optique, CNRS, Universit\'{e} Paris-Saclay, 91127 Palaiseau cedex, France}
\affiliation{SAFRAN Sensing Solutions, Safran Tech, Rue des Jeunes Bois, Ch\^{a}teaufort CS 80112, 78772 Magny-les-Hameaux, France}

\author{Valentin V. Volchkov} 
\affiliation{Laboratoire Charles Fabry, Institut d'Optique, CNRS, Universit\'{e} Paris-Saclay, 91127 Palaiseau cedex, France}
\affiliation{Max-Planck-Institute for Intelligent Systems, Max-Plack-Ring, 4, 72076 T\"{u}bingen, Germany}

\author{Baptiste Lecoutre} 
\affiliation{Laboratoire Charles Fabry, Institut d'Optique, CNRS, Universit\'{e} Paris-Saclay, 91127 Palaiseau cedex, France}

\author{Musawwadah Mukhtar} 
\affiliation{Laboratoire Charles Fabry, Institut d'Optique, CNRS, Universit\'{e} Paris-Saclay, 91127 Palaiseau cedex, France}

\author{Fred Jendrzejewski} 
\affiliation{Laboratoire Charles Fabry, Institut d'Optique, CNRS, Universit\'{e} Paris-Saclay, 91127 Palaiseau cedex, France}
\affiliation{Heidelberg University, Kirchhoff-Institut f\"{u}r Physik, Im Neuenheimer Feld 227, 69120 Heidelberg, Germany}

\author{Alain Aspect} 
\affiliation{Laboratoire Charles Fabry, Institut d'Optique, CNRS, Universit\'{e} Paris-Saclay, 91127 Palaiseau cedex, France}

\author{Adrien Signoles} 
\affiliation{Laboratoire Charles Fabry, Institut d'Optique, CNRS, Universit\'{e} Paris-Saclay, 91127 Palaiseau cedex, France}

\author{Laurent Sanchez-Palencia} 
\affiliation{CPHT, Ecole Polytechnique, CNRS, Universit\'{e} Paris-Saclay, Route de Saclay, 91128 Palaiseau, France}

\author{Vincent Josse}
\email[Corresponding author: ]{vincent.josse@institutoptique.fr} 
\affiliation{Laboratoire Charles Fabry, Institut d'Optique, CNRS, Universit\'{e} Paris-Saclay, 91127 Palaiseau cedex, France}

\date{\today}

\begin{abstract}
We report on an extensive study of the elastic scattering time $\tauS$ of matter-waves in optical disordered potentials. Using direct experimental measurements, numerical simulations and comparison with first-order Born approximation based on the knowledge of the disorder properties, we explore the behavior of $\tauS$ over more than three orders of magnitude, spanning from the weak to the strong scattering regime. We study in detail the location of the crossover and, as a main result, we reveal the strong influence of the disorder statistics, especially on the relevance of the widely used Ioffe-Regel-like criterion $k\lS\sim 1$. While it is found to be relevant for Gaussian-distributed disordered potentials, we observe significant deviations for laser speckle disorders that are commonly used with ultracold atoms. Our results
are crucial for connecting experimental investigation of complex transport phenomena, such as Anderson localization, to microscopic theories.

\end{abstract}

\maketitle


\textit{Introduction.---}  The elastic scattering time $\tauS$, i.e., the mean time between two successive scattering events, is a fundamental time scale to describe wave propagation in disorder, and is thus at the heart of theoretical description of a wide class of physical systems, from light in the atmosphere or in biological tissues to electrons in solid-state systems~\cite{Rammer2004,Akkermans2007}. Furthermore, $\tauS$ is routinely used to characterize the scattering strength via the dimensionless quantity $k \lS$ ($k$: wave number; $\lS=v\tauS$: mean free path, with $v$ the group velocity), which quantifies the number of oscillations of the wave between two scattering events. In this respect, the criterion $k \lS\sim 1$ is widely accepted to set the limit between the weak scattering regime, where perturbative treatments apply, and the strong scattering regime. It coincides with the Ioffe-Regel criterion associated with Anderson localization for point-like scatterers~\cite{Lagendijk2009}.

\begin{figure}[!b]
	\includegraphics[width=0.48\textwidth]{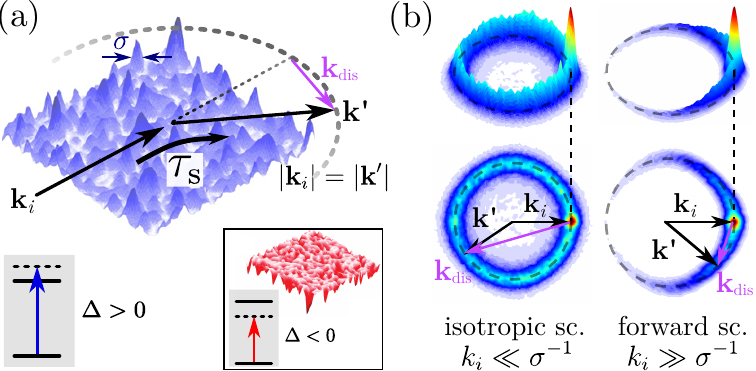}
	\caption{\label{fig:principle} \textbf{Elastic scattering and Born approximation.} (a)~Scattering of a matter-wave by a laser speckle disordered potential of typical correlation length~$\sigma$. During a scattering event, which happens on the characteristic time $\tauS$, a momentum $\kdis$ is transferred to the initial momentum $\ki$. In the Born approximation, the final momentum $\kp=\ki+\kdis$ lies on the elastic scattering ring (dotted circle). For positive atom-light detuning $\Delta>0$, the laser speckle potential is repulsive. Inset : for  $\Delta<0$, it is attractive, having identical spatial properties but opposite amplitude distribution. (b)~Illustrations of the 2D-momentum distributions $n(\textbf{k},t)$ after a typical time $\tauS$ (1st row: side view, 2nd row: top view) for the isotropic ($k_i \ll \sigma^{-1}$) and forward ($k_i  \gg \sigma^{-1}$) scattering regimes.}
\end{figure}

Since $\tauS$ is related to dephasing and not directly to transport properties, its direct determination is rather demanding~\footnote{The transport properties are linked to the transport time $\tau^{\star}$, which is related to the isotropization of the momentum distribution~\cite{Rammer2004,Akkermans2007}.}. So far various measurement methods have been developed, from  Shubnikov-de Haas oscillations of the magneto-conductivity in electronics systems~\cite{Niederer1974,Ando1982,Bockelmann1990,Monteverde2010}, to ballistic transmission~\cite{Page1996,Savo2017}, microscopy techniques~\cite{Jacques2012,Sevrain2013,Martin2016}, and intensity or phase correlations \cite{Shapiro1986,Sebbah2002,Emiliani2003,AnacheMenier2009,Obermann2014,Hildebrand2014} for classical waves. However, the direct comparison between experimental determinations and ab-initio calculations have been scarce (see, e.g.,~\cite{Page1996}) and, to our knowledge, a quantitative investigation of the relevance of the criterion $k \lS\sim 1$ is still lacking. Atomic matter waves in optical disordered potentials offer a controllable platform to investigate the behavior of $\tauS$ with respect to the microscopic details of the disorder. Numerous theoretical predictions exist~\cite{Kuhn2005,Kuhn2007,Skipetrov2008,Hartung2008,Yedjour2010,Cherroret2012,Piraud2012,Shapiro2012,Plisson2013,Piraud2013,Piraud2014}, yielding in particular to the derivation of an alternative condition to the $k \lS\sim 1$ criterion~\footnote{$k \lS\sim 1$ deviates significantly, in the forward scattering regime, from the criterion $\epsilon_k \sim \VR^2/E_\sigma$ (with $E_\sigma=\hbar^2/m\sigma^2$ the correlation energy and $\sigma$ the correlation length of the disoder).} in Ref.~\onlinecite{Kuhn2007}, rendering a precise investigation highly desirable.

In this letter, we report direct measurements of the elastic scattering time $\tauS$ for ultracold atoms propagating in quasi two-dimensional laser speckle disordered potential. The scattering time is directly measured by monitoring the time evolution of the momentum distribution for a wave packet having a well defined initial momentum, and the results are compared to numerical simulations, yielding to an excellent agreement. The simulations are also used to extend our investigation to the case of a Gaussian disorder, a model widely considered in condensed matter physics. Most importantly, we study the evolution of $\tauS$ over a large parameter range ($\tauS$ varies by more than three order of magnitude), allowing us to span from the weak to the strong scattering regime. Comparing our results to analytical 1st-order Born calculations, we reveal the strong influence of disorder statistics on the crossover and discuss the relevance of the Ioffe-Regel-like criterion $k \lS\sim 1$. 


\textit{First-order Born approximation.---} For weak disorder, we can develop an intuitive, physical picture of the scattering time based on the 1st-order Born approximation (referred to as Born approximation in the following)~\cite{Rammer2004,Akkermans2007}. In this perturbative treatment, $\tauS$ can be interpreted as the finite lifetime of the incoming free state $|\ki \rangle $, as it is scattered towards a continuum of final momenta $\ket{\kp}$ with $|\kp|=|\ki|$. The initial momentum distribution therefore decays exponentially in this regime, with the characteristic time $\tauS$:
\begin{equation}
\label{eq:exponential_decay}
n(\ki,t)=n(\ki,0)\,e^{-t/\tauS} \ ,
\end{equation} 
where $t$ is the propagation time in the disorder. The scattering is only allowed if there exists a spatial frequency component $\kdis$ in the disordered potential that matches the elastic scattering condition $\kdis=\kp-\ki$ [Fig.~\ref{fig:principle}(a)]. The weight of scattering in this direction relies uniquely on the spatial frequency distribution of the disorder $\tilde{C}(\kdis)$, i.e., the Fourier transform of the two-point correlation function $C(\Delta \mathbf{r})=\overline{ V(\mathbf{r})V(\mathbf{r} +\Delta \mathbf{r})  }$ (where $\overline{\cdots}$ refers to disorder averaging). Using the Fermi golden rule, the Born elastic scattering time $\tauSB$ is obtained by summing the contributions coming from the scattering in all directions, yielding: 
\begin{equation}
\label{eq:fermi_golden_rule}
\frac{\hbar}{\tauSB}= 2\pi\,\sum_{\kp}\tilde{C}(\kp-\ki)\  \delta[ \epsilon_{k'}-\epsilon_{k_i}  ] \ ,
\end{equation}
where $\epsilon_k=\hbar^2 k^2 /2 m$ is the free-state energy, with $m$ the atomic mass. 

\begin{figure*}[!t]
\includegraphics[width=0.99\textwidth]{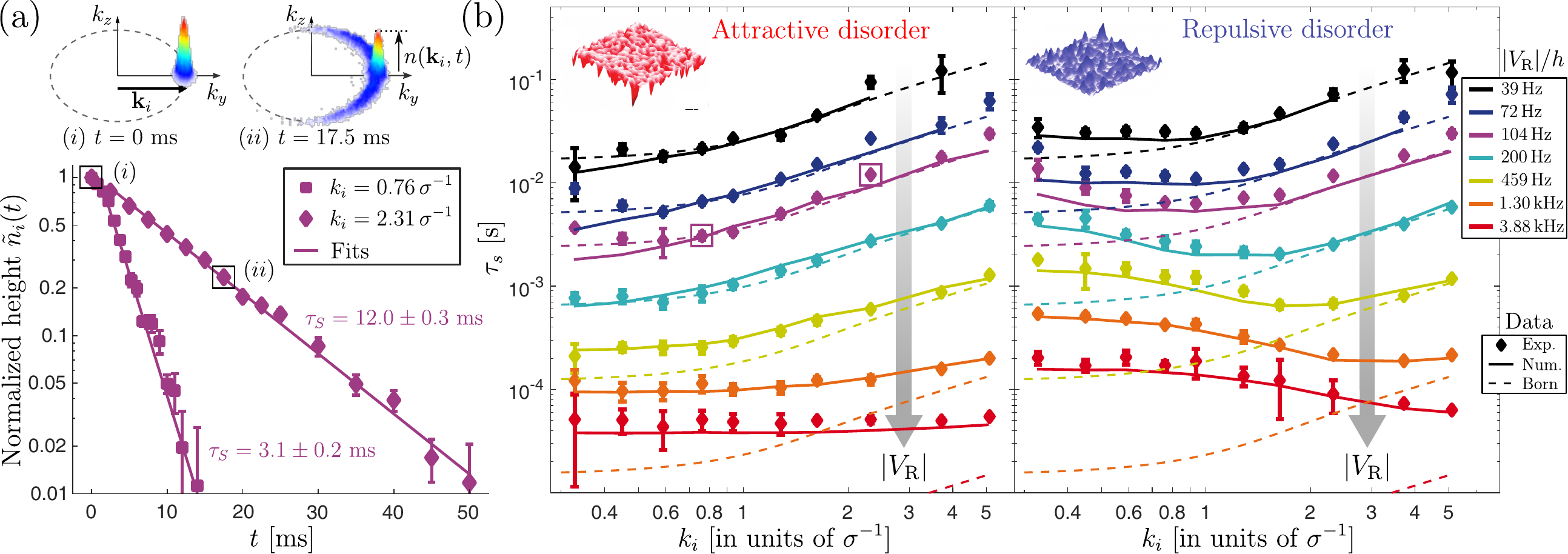}
\caption{\label{fig:mainresult} \textbf{Measurements of the elastic scattering time $\mathbf{\tauS}$.} 
(a) Measurement procedure. The momentum distributions $n(\textbf{k},t)$ are observed for different propagation times $t$ in the disorder, here shown for the parameters $\VR/h=-104$~Hz (attractive case) and $k_i=2.31\, \sigma^{-1}$. The normalized height $\tilde{n}_i(t) $ is determined from $n(\textbf{k},t)$ by a Gaussian fit of the radially integrated angular profile~\cite{SM}. When plotted as a function of the time $t$, it shows an exponential decay from which we extract $\tauS$, as illustrated for two different initial momenta $k_i=0.76\, \sigma^{-1}$ and  $k_i=2.31\, \sigma^{-1}$, still at $\VR/h=-104$~Hz.
(b) Experimental (points) and numerical (solid lines) values of $\tauS$ as function of the initial momentum $k_i$ for different values of the disorder amplitude $|\VR|$, in the cases of an attractive disorder (left panel) and a repulsive disorder (right panel). The initial momenta are shown in units of the characteristic frequency $\sperp^{-1}$ of the disorder. Born predictions $\tauSB$ are indicated by dashed lines. Note that the Born curves are simply shifted down for the various disorder amplitudes due to the scaling $\tauSB\propto1/|\VR|^2$ (see text).  }
\end{figure*}

The correlation length $\sigma$ of the disorder, i.e., the typical width of $C(\Delta \mathbf{r})$, introduces a characteristic spatial frequency $\sigma^{-1}$ that defines two scattering regimes. For low initial momentum $k_i\ll \sigma^{-1}$, the disorder contains the spatial frequencies that are necessary to scatter the atoms in all directions and the scattering is isotropic [see Fig.~\ref{fig:principle}(b)]. In the opposite case of large momentum $k_i \gg \sigma^{-1}$, the disorder's spatial frequencies are too small for satisfying the backward scattering condition ($\kdis=-2\ki$) and the scattering is essentially concentrated in the forward direction. As discussed in~\cite{Kuhn2007,Piraud2012,Piraud2013,Shapiro2012}, the Born prediction~\eqref{eq:fermi_golden_rule} yields different behaviors in the two regimes: $\tauSB$ is essentially constant for isotropic scattering while it increases linearly with momentum in the forward case (see dashed lines in Fig.~\ref{fig:mainresult} and~\cite{SM} for further details). 

Note that the validity of the Born approximation can be estimated in an intuitive manner. Due to its finite lifetime $\tauSB$, the matter wave acquires a finite energy width $\Delta \epsilon =\hbar /\tauSB$ [responsible for the ring's width seen in Fig.~\ref{fig:principle}(b)]. By consistency, $\Delta \epsilon$ should be much smaller than the initial energy $ \epsilon_{k_i}\propto k_i^2 $, yielding  the usual weak scattering criterion $k_i\lSB \gg 1$ introduced above (with $\lSB \propto k_i \tauSB$). In the following we study experimentally and numerically the validity of this criterion by analyzing scattering times for various potential disorders $V(\mathbf{r})$ and over a large range of initial momentum $\ki$, allowing us to investigate the crossover between weak and strong scattering. 


\textit{Experiment.---}  Based on Eq.~\eqref{eq:exponential_decay}, we directly measure $\tauS$ by monitoring the decay of the initial momentum distribution of atoms launched with a well defined initial momentum $\ki$ into a disordered potential $V(\mathbf{r})$~\cite{Cherroret2012}. The experimental set-up is similar to the one described in Refs.~\cite{Jendrzejewski2012CBS,Muller2015}. It relies on the production of a quasi non-interacting cloud of $10^5$ $^{87}$Rb atoms in a $F=2,\,m_F=-2$ Zeeman sublevel, suspended against gravity by a magnetic field gradient. A delta-kick cooling sequence leads to an ultra-narrow momentum spread $\Delta k = \SI{0.15}{\micro\meter^{-1}}$ ($T\sim 150$~pK). A mean initial momentum $\textbf{k}_{i}$, ranging from $k_{i} = \SI{1}{\micro\meter^{-1}}$ to $k_{i} = \SI{20}{\micro\meter^{-1}}$ along the $y$ axis, is then given to the atoms by pulsing an external magnetic gradient for a tunable duration. 

\begin{figure}[!t]
\includegraphics[width=0.49\textwidth]{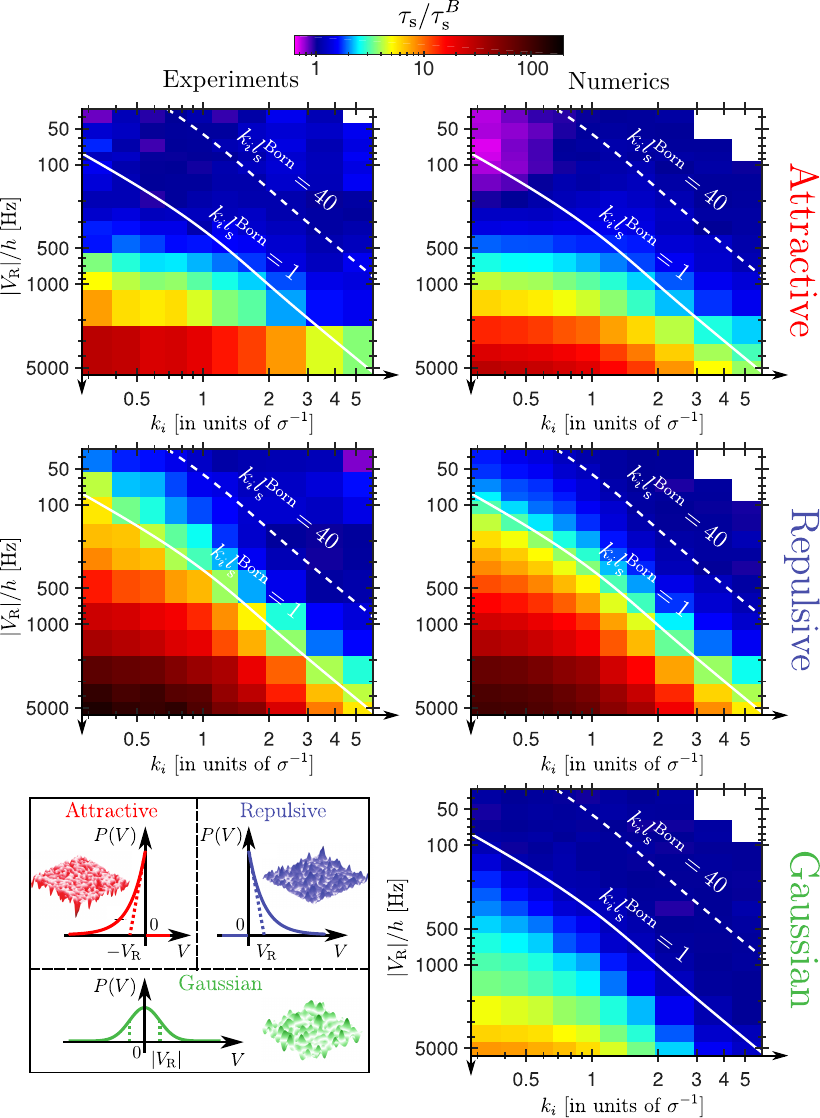}
\caption{\label{fig:Deviations} \textbf{Deviations from the Born predictions for different disorder amplitude distributions.} 2D representation (logarithmic color scale) of the ratio $\tauS /\tauSB$ as a function of $|\VR|$ and $k_i$ for attractive (1st row) and repulsive (2nd row) disordered potentials. Both experimental (left column) and numerical (right column) data are shown. 3rd row: same representation for a Gaussian-distributed disorder (numerical study). The amplitude probability distributions $P(V)$ for the three types of disorders are plotted in the inset.}
\end{figure}

A quasi-2D disordered potential in the $(y-z)$ plane is created by a laser speckle field~\cite{Clement2006,Goodman2007}, realized by passing a laser beam along the $x$ axis through a rough plate and focusing it on the atoms~\cite{SM}. The wavelength of the laser is red- or blue-detuned with respect to the atomic transition ($D_2$ line of $^{87}$Rb around $\SI{780}{\nano \meter}$) in order to create either an attractive or a repulsive disordered potential [see Fig.~\ref{fig:principle}(a)]. The detuning being small enough ($\Delta\sim \SI{1}{\tera \hertz}$), both disorders have the same spatial correlation function, with a measured correlation length $\sperp=\SI{0.50(1)}{\micro m}$ (radius at $1/e$). 
However, they differ by their probability distribution $P(V)$, both exhibiting the asymmetrical exponential distribution of laser speckle fields~\cite{Goodman2007}, but with opposite signs (see inset of Fig.~\ref{fig:Deviations}): $P(V)= | \VR |^{-1} e^{-V / \VR} \cdot \Theta (V/ \VR ) $, with  $\Theta$ the step function. Here $\VR$ is the averaged amplitude (negative for attractive and positive for repulsive laser speckle), while the rms disorder amplitude, i.e., the quantity that characterizes the disorder strength, is the absolute value $|\VR|$. When varying the laser power and detuning, $|\VR|/h$ ranges from $\SI{39}{Hz}$ to $\SI{3.88}{kHz}$.

The experimental sequence starts with the preparation of an atomic cloud with momentum $\ki$. At $t=0$ we rapidly switch on the disorder potential $V(\mathbf{r})$, performing a quantum quench of the system. After a time evolution $t$, the disorder is switched off and we record the momentum distribution $n(\textbf{k},t)$ by fluorescence imaging after a long time of flight. Thanks to gravity compensation, up to $\SI{300}{\milli \second}$ can be achieved, corresponding to a momentum resolution $\Delta k_\mathrm{res}=\SI{0.2 }{\micro \meter}^{-1}$~\footnote{It includes the initial momentum spread $\Delta k$ and the initial size of the cloud of $\SI{30}{\micro \meter}$.}. From these images we extract the evolution of the initial momentum population $n(\ki,t)$, as shown in Fig.~\ref{fig:mainresult}(a)~\cite{SM}. At low disorder strength $|\VR|$, an exponential decay is observed for almost two orders of magnitude and a fit yields the experimental value of $\tauS$ [refer to Eq.~\eqref{eq:exponential_decay}]. 
Such exponential decay is expected to persist at larger disorder amplitudes, except if one drives the system to the very strong scattering regime (see e.g.~\cite{Trappe2015}). However, no significant departure from an exponential decay was observed in our experiment and all the recorded decays could be fitted by an exponential function.

\textit{General results.---} Figure~\ref{fig:mainresult}(b) shows the measured values of the elastic scattering time $\tauS$ for both the attractive and repulsive laser speckle disorder cases. The large set of disorder amplitude and initial momenta  
allows us to observe variations of $\tauS$ from $\SI{40}{\micro\second}$ to $\SI{100}{\milli\second}$. These observations are compared to 2D numerical calculations (solid lines)~\cite{SM}, with a remarkable agreement over almost the whole data range, confirming the quasi-2D character of our configuration. Deviations are nevertheless observed in a small zone (very low momenta and disorder amplitudes, upper left part on the graphs) and may be attributed to technical difficulties to precisely measure $\tauS$ in this regime due to the finite momentum resolution $\Delta k_\mathrm{res}$.

The Born prediction~\eqref{eq:fermi_golden_rule} is also shown in Fig.~\ref{fig:mainresult}(b) (dashed lines)~\footnote{It corresponds to 2D calculations, which are found in excellent agreement with full 3D calculations~\cite{SM}}. Note that $\tauSB$ scales with the rms value $|\VR|$ as $1/|\VR|^2$~\cite{SM}, but does not depend on the specific form of the disorder amplitude distribution $P(V)$. As a consequence, the prediction is strictly identical for both attractive and repulsive speckles, since they possess the same frequency distribution $\tilde{C}(\kdis)$. In general, $\tauSB$ shows a very good agreement with the data at low scattering strength, i.e., weak $|\VR|$ and large $k_i$ [upper right part on Fig.~\ref{fig:mainresult}(b)], as expected for this first order perturbative approach. However significant deviations appear at the lowest disorder amplitude ($|\VR|/h=39$~Hz, black dots) when considering the low initial momentum range $k_i\lesssim \sigma^{-1} $. As the disorder strength $|\VR|$ increases, the deviations become more pronounced and extend to larger momenta. In strong scattering conditions, the two regimes previously identified (isotropic and forward scattering) are then not relevant anymore. Moreover, large differences are observed between attractive and repulsive disorders, another signature of the complete failure of the Born approximation.

In order to visualize these deviations, we show in Fig.~\ref{fig:Deviations} maps of the ratio $\tauS / \tauSB$ as a function of the parameters $k_i$ and $|\VR|$. The important role of the disorder statistics is further emphasized by numerically extending our analysis to the case of a disorder with a Gaussian amplitude probability distribution $P(V)=(\sqrt{2\pi}\VR)^{-1} e^{-V^2/2\VR^2}$ (inset of Fig.~\ref{fig:Deviations}), $|\VR|$ being still the rms value. For consistency, we have chosen the same two-point correlation function $C(\Delta \mathbf{r})$ as the one of the laser speckles considered so far~\cite{SM}. 


\textit{Weak to strong scattering crossover.---}  The maps shown in Fig.~\ref{fig:Deviations} allow us to investigate the crossover between the weak (Born regime) and strong scattering regimes. Considering first the case of a Gaussian-distributed disorder (3rd row), we observe a striking coincidence between the iso-deviation lines and the dimensionless parameter $k_i \lSB $. In particular, the $k_i \lSB= 1$ line, i.e. the usual criterion introduced earlier, corresponds to a typical deviation of 25$\%$. Importantly enough, this observation confirms, in a quantitative manner, the relevance of the criterion $k_i \lSB = 1 $ to differentiate the weak and strong scattering regimes for this commonly used type of disorder.

In contrast, this criterion does not hold for laser speckle disorders, for which the deviations to the Born prediction are much more pronounced. For instance, the $k_i \lSB= 1$ line corresponds now to deviations up to $250\%$ for the attractive case (1st row) and to $400\%$ for the repulsive case (2nd row). As a result, the crossover is significantly shifted towards larger $k_i \lSB$ values, i.e. larger momenta and lower disorder amplitudes. More precisely, the same 25$\%$ deviation as considered above corresponds to an effective criterion $k_i \lSB=40$ (white dashed lines).


\textit{Beyond the 1st-order Born approximation.---} An exhaustive description of the deviations from the Born prediction is beyond the scope of the present letter~\cite{LongPaperTaus}. It is however possible to get some physical insight by considering two different regimes. First, in the intermediate scattering regime of low momenta and low disorder amplitude (upper left part of the maps in Fig.~\ref{fig:Deviations}), the deviations can still be understood within perturbative theory~\cite{Rammer2004,Akkermans2007}, going to higher order corrections~\cite{Kuhn2007,Lugan2009,Gurevich2009}. Since the next higher-order term scales as $1/\VR^3$, it is negative for attractive speckle disorder, positive for the repulsive one, but vanishes for Gaussian disorder due to the symmetry of the probability distribution. This explains the important difference between the three types of disorder in this parameter range.

When going to the very strong scattering regime (lower left part of the maps), the perturbative approach completely breaks down. To interpret the data, it is then fruitful to invoke the general concept of spectral functions $A_{\ki}(E)$, which give the energy probability distribution of the initial state $\ket{\ki}$ once the disorder is suddenly switched on. Their width is indeed inversely proportional to the measured scattering time $\tauS$~\cite{bruus}. In the strong disorder limit, the spectral functions are known to converge towards the disorder amplitude distribution $P(V)$~\cite{Trappe2015,Prat2016,Volchkov2018}. As a result $\tauS$ essentially scales as $1/|\VR|$ in this limit, yielding values well above the Born prediction (scaling as $1/|\VR|^2$, see above). That general trend explains the large positive deviations observed in Fig.~\ref{fig:Deviations}. In this regime, the specific shape of the spectral functions associated to each type of disorder leads however to discrepancies for the measured scattering times~\cite{Trappe2015,Prat2016,Volchkov2018}. In particular the spectral functions for the repulsive speckle disorder exhibit a narrow peak at low energy~\footnote{This peak can be traced to the presence of many bound states, with similar energies, that are supported by the valleys formed around local minima of the potential~\cite{Trappe2015,Prat2016,Volchkov2018}.}, which is responsible for the striking increase of the scattering time (almost two orders of magnitude from the Born prediction). In order to support this analysis, we have verified the very good agreement between the present measurements and the width of the spectral functions recently measured for laser speckle disorders in Ref.~\onlinecite{Volchkov2018}.


\textit{Conclusion.---} Combining direct experimental measurements, numerical simulations and comparison with ab-initio Born calculations, we have provided an extensive analysis of the elastic scattering time $\tauS$ for ultracold atoms in disordered potential. Using the large accessible range of parameters, we have demonstrated the strong influence of the disorder statistics on the relevance of the commonly accepted $k\lS \sim 1$ criterion to identify the crossover from the weak to strong scattering regime: while it is relevant for Gaussian disorder, large deviations are reported for laser speckle disorder.

Our results open various prospects. On the theory side, a natural follow up would be to go beyond the Born approximation and compare our data to higher order perturbative treatments~\cite{Kuhn2007,Lugan2009,Gurevich2009,Piraud2014}, self-consistent Born approximation~\cite{Vollhardt1980,Skipetrov2008,Yedjour2010}, the recently developed Schwinger-Ward-Dyson theory~\cite{Sbierski2018}, or semiclassical approaches~\cite{Trappe2015,Prat2016}. On the experimental side, the precise knowledge of the elastic scattering time for laser speckle reported here is of particular importance in view, for instance, to investigate Anderson localization~\cite{Jendrzejewski2012,Semeghini2015}. This work then paves the way for further experimental investigation in strong connection with microscopic theories, either using the spectroscopic scheme proposed in Ref.~\onlinecite{Volchkov2018}, or searching for direct signatures in momentum space~\cite{Cherroret2012,Jendrzejewski2012CBS,Muller2015,Karpiuk2012,Ghosh2017,Hainaut2018}.

\begin{acknowledgments}
We would like to thank D. Delande, T. Giamarchi, G. Montambaux, C. M\"uller and S. Skipetrov  for fruitful discussions and comments. This work has been supported by ERC (Advanced Grant ``Quantatop''), Institut Universitaire de France, the French Ministry of Research and Technology (ANRT, through a DGA grant for J.~R. and CIFRE/DGA grant for V.~D.), the EU-H2020 research and innovation program (Grant No. 641122-QUIC and Marie Sk\l odowska-Curie Grant No. 655933), Region Ile-de-France in the framework of DIM SIRTEQ and the Simons foundation (award number 563916: Localization of waves).
\end{acknowledgments}


\begin{thebibliography}{51}
\bibitem [{Lon()}]{LongPaperTaus}%
  \BibitemOpen
  \href@noop {} {}\bibinfo {note} {A detailed analysis is in
  preparation.}\BibitemShut {Stop}%
\end{thebibliography}
%

\end{document}



\title{Supplemental Material: Elastic Scattering Time of Matter-Waves in Disordered Potentials}

\author{J\'{e}r\'{e}mie Richard} 
\affiliation{Laboratoire Charles Fabry, Institut d'Optique, CNRS, Universit\'{e} Paris-Saclay, 91127 Palaiseau cedex, France}

\author{Lih-King Lim} 
\affiliation{Zhejiang Institute of Modern Physics, Zhejiang University, Hangzhou 310027, P. R. China}
\affiliation{Laboratoire Charles Fabry, Institut d'Optique, CNRS, Universit\'{e} Paris-Saclay, 91127 Palaiseau cedex, France}

\author{Vincent Denechaud} 
\affiliation{Laboratoire Charles Fabry, Institut d'Optique, CNRS, Universit\'{e} Paris-Saclay, 91127 Palaiseau cedex, France}
\affiliation{SAFRAN Sensing Solutions, Safran Tech, Rue des Jeunes Bois, Ch\^{a}teaufort CS 80112, 78772 Magny-les-Hameaux, France}

\author{Valentin V. Volchkov} 
\affiliation{Laboratoire Charles Fabry, Institut d'Optique, CNRS, Universit\'{e} Paris-Saclay, 91127 Palaiseau cedex, France}
\affiliation{Max-Planck-Institute for Intelligent Systems, Max-Plack-Ring, 4, 72076 T\"{u}bingen, Germany}

\author{Baptiste Lecoutre} 
\affiliation{Laboratoire Charles Fabry, Institut d'Optique, CNRS, Universit\'{e} Paris-Saclay, 91127 Palaiseau cedex, France}

\author{Musawwadah Mukhtar} 
\affiliation{Laboratoire Charles Fabry, Institut d'Optique, CNRS, Universit\'{e} Paris-Saclay, 91127 Palaiseau cedex, France}

\author{Fred Jendrzejewski} 
\affiliation{Laboratoire Charles Fabry, Institut d'Optique, CNRS, Universit\'{e} Paris-Saclay, 91127 Palaiseau cedex, France}
\affiliation{Heidelberg University, Kirchhoff-Institut f\"{u}r Physik, Im Neuenheimer Feld 227, 69120 Heidelberg, Germany}

\author{Alain Aspect} 
\affiliation{Laboratoire Charles Fabry, Institut d'Optique, CNRS, Universit\'{e} Paris-Saclay, 91127 Palaiseau cedex, France}

\author{Adrien Signoles} 
\affiliation{Laboratoire Charles Fabry, Institut d'Optique, CNRS, Universit\'{e} Paris-Saclay, 91127 Palaiseau cedex, France}

\author{Laurent Sanchez-Palencia} 
\affiliation{CPHT, Ecole Polytechnique, CNRS, Universit\'{e} Paris-Saclay, Route de Saclay, 91128 Palaiseau, France}

\author{Vincent Josse}
\email[Corresponding author: ]{vincent.josse@institutoptique.fr} 
\affiliation{Laboratoire Charles Fabry, Institut d'Optique, CNRS, Universit\'{e} Paris-Saclay, 91127 Palaiseau cedex, France}


\begin{abstract}
We describe here the methods (i) to extract the scattering time from the measurement, (ii)  to generate and characterize the laser speckle field, (iii) to calculate the Born prediction adapted to our configuration and (iv) to perform the numerical simulations to estimate the scattering times.
\end{abstract}

\maketitle

\section{Extraction of the scattering time $\tauS$}

The measurement of the elastic scattering time $\tauS$ is based on the decay of the initial momentum distribution $n(\ki,t )$, as illustrated in Fig.~\ref{fig:tauS_extraction} for the parameters $k_i=2.31 \sigma^{-1}$ and $\VR/h=\SI{-104}{\hertz}$ (same as in Fig.2(a) on the main text).

The determination of $\tauS$ is performed in three steps. 
First, we determine the angular profile $n(\theta, t)$ by radially integrating the momentum distribution $n(\mathbf{k},t)$ in the ($k_y,\,k_z$) plane, $\theta=0$ corresponding to the initial direction [Fig.~\ref{fig:tauS_extraction}(a)]. The lower and upper integration limits correspond to twice the radial width of the initial momentum distribution. The reduced angular profile $\tilde{n}(\theta,t)=n(\theta, t)/n(0,0)$ is obtained by normalizing this profile by its initial value at time $t=0$ and $\theta=0$. A typical angular profile $\tilde{n}(\theta,t)$ for $t=\SI{17.5}{\milli\second}$ is plotted on~Fig.{\ref{fig:tauS_extraction}(b) (blue line). The general shape results from the sum of two contributions: a narrow peak $\nin(\theta,t)$ that corresponds to the unscattered initial distribution and a broad background $\nb(\theta,t)$. The latter corresponds to the scattered atoms to direction $\kp$ and builds up progressively on time. 

In a second step, the normalized height $\nin(t)=\nin(0,t)$ of the initial distribution is extracted by adjusting the bi-modal distribution by the sum of a narrow Gaussian peak accounting for $\nin(\theta,t)$~\footnote{The initial momentum is not an inverted parabola ---which is expected for a Bose Einstein condensate in the Thomas Fermi regime--- but merely resembles to a Gaussian once the delta-kick cooling technique has been applied.} and a broad Gaussian peak accounting for the background $\nb(\theta,t)$ [red solid line in Fig.\ref{fig:tauS_extraction}(b)]. Error bars of $\nin(t)$ represent one standard deviations and are estimated from the noise on the experimental data and the deviation of the model. 

To finally extract $\tauS$, the decay of $\nin(t)$ is plotted in a semi-logarithmic scale [dots in Fig.~\ref{fig:tauS_extraction}(c)] and then adjusted by a pure exponential law of typical time $\tauS$.

\begin{figure}[!b]
\includegraphics[width=0.49\textwidth]{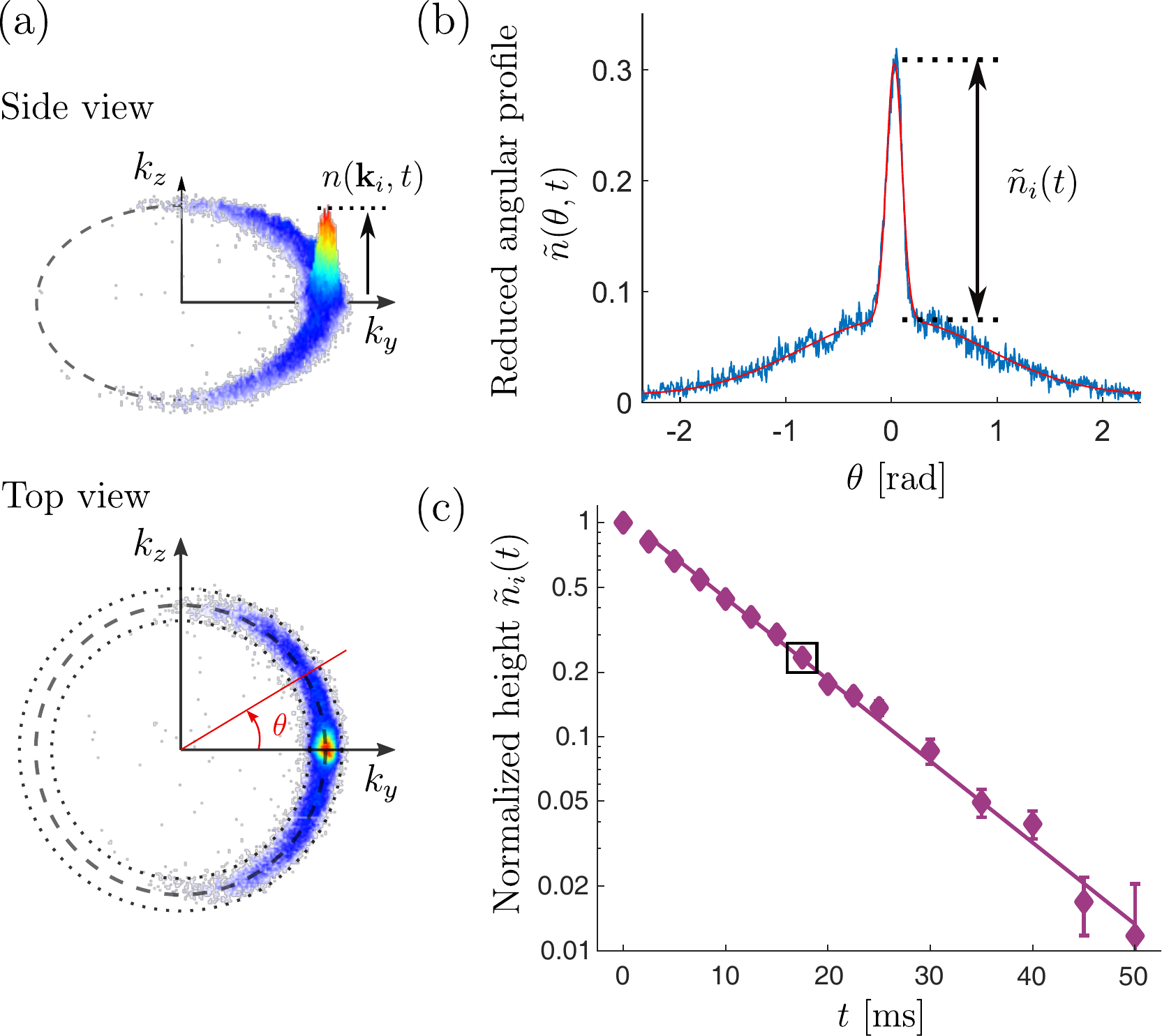}
\caption{\label{fig:tauS_extraction} \textbf{Measurement protocol of the scattering time $\mathbf{\tauS}$.} The procedure is illustrated for the same parameters than in Fig.~2(a) of the main text, i.e., $k_i=2.31 \sigma^{-1}$ and $\VR=\SI{-104}{\hertz}$. (a) Observed momentum distribution $n(\mathbf{k},t)$ after a propagation time $t=\SI{17.5}{\milli\second}$ in the disorder (top: side view; bottom: top view). The angular profile is obtained by radially integrating around the initial peak (between the two dotted lines). (b) The reduced angular profile $\tilde{n}(\theta,t)$ (blue line) is adjusted by the sum of a narrow and a broad Gaussian peak, both centered around $\theta=0$ (red line). The amplitude of the narrow peak is used to extract the normalized height $\nin(t)$. (c) The normalized height $\nin(t)$ is plotted as a function of the propagation time $t$. The experimental points are fitted by an exponential function of the form $e^{-t/\tauS}$. Here we find $\tauS = \SI{12.0 \pm 0.3}{\milli\second}$.}
\end{figure}
  
\section{Laser speckle disordered potential}

\subsection{Quasi-2D laser speckle field generation}

The laser speckle field is created by passing a laser beam of wavelength $\lambda \sim 780$~nm through a diffusive plate, the configuration being identical as the one described in Ref.~\cite{Volchkov2018}. As illustrated in Fig.~\ref{fig:corr}(a), the incoming wave that illuminates the diffusive plate is converging at the position $d=15.2(5) $~mm that coincides with the position of the atoms. The intensity profile of the illumination on the diffusive plate is a Gaussian shape, of waist $w=9(1)$~mm (radius at $1/e^2$), truncated by a circular diaphragm of diameter $D=20.3(1)$~mm. This diaphragm sets the maximal numerical aperture to $\mathrm{NA}=\sin(\theta_\mathrm{max})=0.55(2)$.

In this configuration a so-called Fourier speckle pattern is formed around the position of the atoms. In order to characterize it, the random intensity pattern was recorded at the position of the atoms with an high-resolution optical microscope, see Fig.~\ref{fig:corr}(b). As can be seen, the laser speckle field is very elongated along the propagation axis ($x$ direction), resulting in a quasi-2D potential.

\begin{figure}[b!]
\includegraphics[width=1\columnwidth]{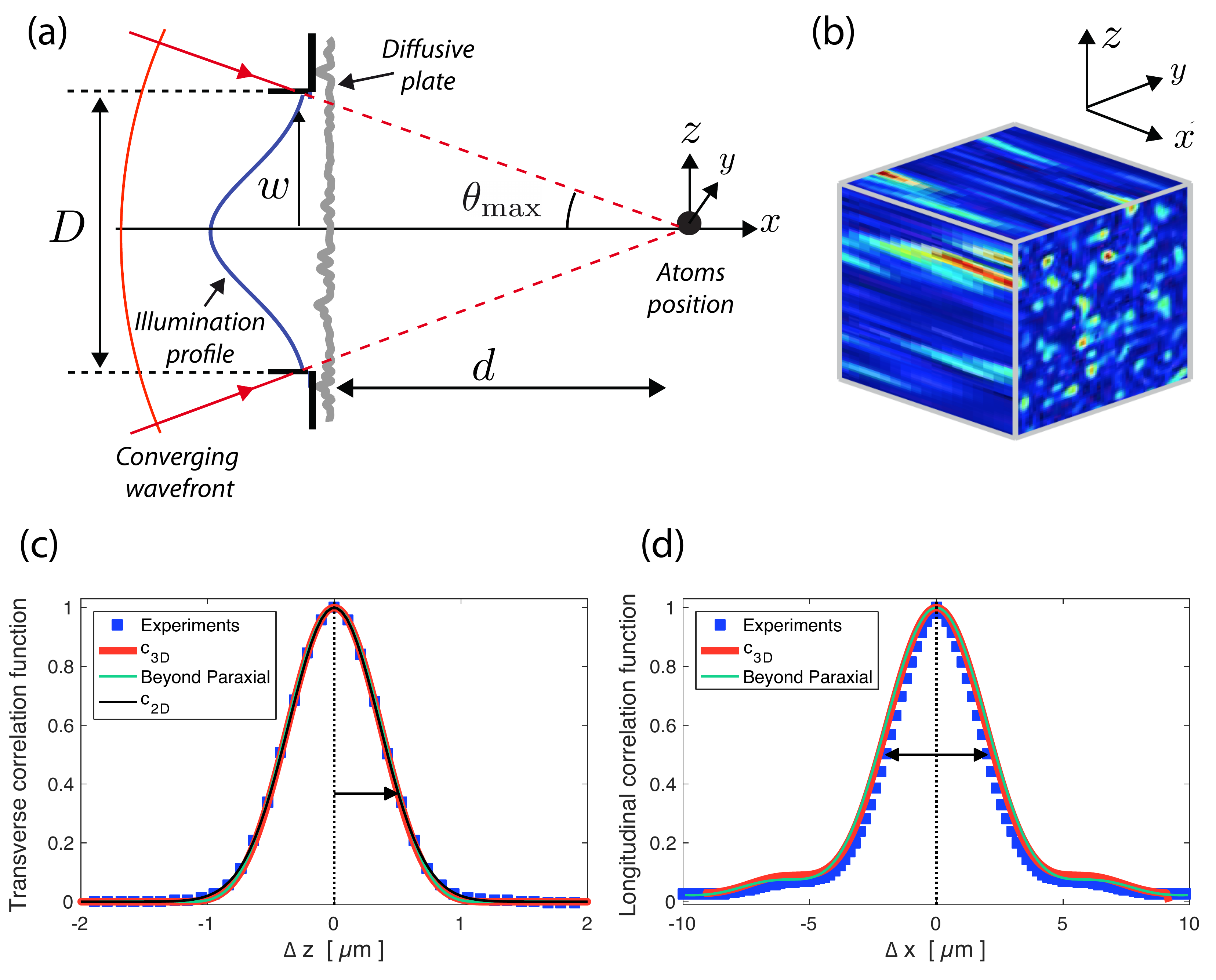}
\caption{\textbf{Laser speckle characterization.} (a) Schematic representation of the experimental configuration used for the speckle generation, with $\mathrm{NA}=\sin(\theta_\mathrm{max})=0.55(2)$. (b) 3D representation of the experimental laser speckle field. (c) Transverse correlation function along the $z$ direction (same along $y$). Blue squares: experimental measurement $c_\mathrm{exp}$ from the recorded intensity pattern. Black solid line: transverse Gaussian fit $\CdeuxD$, yielding $\sigma=\SI{0.50(1)}{\micro \meter}$ ($1/e$ radius). Red solid line: effective paraxial calculations $\CtroisD$ (see text). Green solid line: beyond paraxial calculations, as done in Ref.~\onlinecite{Volchkov2018}. This line is barely visible due to superposition with the effective paraxial model. (d) Longitudinal correlation functions with the same legend as in $(c)$. The measured width is $\sigma_\|=\SI{4.1(1)}{\micro \meter}$ (FWHM) . }
\label{fig:corr}
\end{figure}

\subsection{Spatial statistical properties: measurement of the auto-correlation function}

The normalized two-point correlation function of the laser speckle field,
\begin{equation}
c_\mathrm{exp}(\Delta \mathbf{r})=\frac{\langle \delta I(\mathbf{r}) \; \delta I(\mathbf{r} + \Delta \mathbf{r})  \rangle }{ \langle \delta I^2 \rangle} \quad \mbox{with} \;\,  \delta I=I-\langle I \rangle\,,
\label{Eq:Corr_norm}
\end{equation}
is directly calculated from the recorded spatial pattern shown in Fig.~\ref{fig:corr}(b). The resulting transverse and longitudinal correlation functions are respectively shown as blue squares in Fig.~\ref{fig:corr}(c)~and~(d).

 In the transverse plane, the shape is found to be very close to a Gaussian. A fit of the form (solid black line):
 \begin{equation}
 \CdeuxD (\Delta \mathbf{r}_\perp)= e^{-\Delta \mathbf{r}_\perp^2/\sigma^2} \; ,
 \label{Eq:Corr_2DGauss}
 \end{equation}
realized either along the $y$ or $z$ axis, yields $\sigma=\SI{0.50(1)}{\micro \meter}$ ($1/e$ radius). 

The laser speckle field being very elongated, the correlation function has a much larger width in the longitudinal direction $x$. It is characterized by the FWHM $\sigma_\|=\SI{4.1(1)}{\micro \meter}$ [see Fig.~\ref{fig:corr}(d)].  
  
 \subsection{Modeling the laser speckle field}
 
Due to the large numerical aperture $\mathrm{NA}=\sin(\theta_\mathrm{max})=0.55(2)$, the precise modeling of the speckle field requires in principle to go beyond the paraxial approximation. A theoretical model (not detailed here) was thus used in Ref.~\onlinecite{Volchkov2018} to reproduce the measured correlation functions, see green solid lines in Fig.~\ref{fig:corr}(c) and (d). 

 However this theoretical model is quite heavy to handle, especially in view of the determination of the 3D spatial frequency distribution $\tilde{C}(\kdis)$, a key quantity to calculate the Born prediction. Thus, we developed a simpler model, based on the paraxial approximation but including a global geometrical factor $x_\mathrm{scale}$ to tune the numerical aperture. In this effective paraxial model, the correlation function can be calculated using Fourier Transform (FT)~\cite{Goodman2007}:
 \begin{equation}
 \CtroisD(\Delta \mathbf{r}_\perp,\Delta x)\propto \Big| \, \mathrm{FT}\Big[ \; t(\mathbf{R}_\perp)\; I_\mathrm{inc} (\mathbf{R}_\perp) \; e^{-i \pi \frac{R_\perp^2 \Delta x}{\lambda d} }\Big]_{\frac{\Delta  \mathbf{r}_\perp}{\lambda d}} \, 
 \Big|^2 ,
 \label{Eq:CorrParaxial}
 \end{equation}
where $ \mathrm{FT}\,\big[\,f(x)\,\big]_u=\int_{-\infty}^{+\infty} dx\, f(x) e^{-2i\pi u x}$. Here $t=\mathrm{disc}\big[R_\perp/(D_\mathrm{eff})\big]$ represent the transmission of a circular diaphragm of diameter $D_\mathrm{eff}=x_\mathrm{scale} \, D$, and $I_\mathrm{inc}=e^{-2 R_\perp^2/(w_\mathrm{eff})^2}$ is the Gaussian illumination profile on the diffuser, with effective waist $w_\mathrm{eff}=x_\mathrm{scale} \, w$. 
 
Setting $x_\mathrm{scale} =0.875(5)$ (resulting in an effective maximal numerical aperture $\mathrm{NA}_\mathrm{eff}=0.5$), the calculated correlation function $\CtroisD(\Delta \mathbf{r})$ matches also very well with the measurements, both on the transverse and longitudinal directions [see red solid lines  in Fig.~\ref{fig:corr}(c) and (d)]. Thus, we used this effective paraxial model to calculate the Born prediction for our specific disorder configuration (see below). 

\subsection{Calibration of the disorder amplitude $\VR$}

 In  practice, the disorder amplitude $\VR$ can be calibrated by combining photometric measurements and calculation of the atomic polarizability~\cite{Clement2006}. However it is known that such method leads to systematic uncertainties, typically around a few tens of percents (see e.g. Refs.~\onlinecite{Semeghini2015,Volchkov2018}). 
 
Here, we used the excellent agreement between the experimental  determination of $\tauS$ and the numerical simulations [see Fig.~2(b) of the main text] to precisely determine the disorder amplitude by applying an overall correction $\alpha$ on the photometric measurement. In practice, the correction factor is calculated by minimizing the differences between the experiments and numerics for the particular momenta $k_i=0.74\, \sigma^{-1}$, leading to $\alpha=1.29(2)$. We mainly attribute this correction to the difficulty to estimate precisely the extension of the speckle field at the position of the atoms.
 
\section{First order Born approximation}\label{sec:Born}

\subsection{Rescaled scattering times $\tauSBtilde$}

An important feature of the Born prediction is the simple scaling $\tauSB\propto1/|\VR|^2$ with the disorder amplitude. Indeed, the spatial frequency distribution of the disorder can be written in the form $\tilde{C}(\kdis)=|\VR|^2 \tilde{c}(\kdis)$, where $\tilde{c}(\kdis)$ is the Fourier transform of the normalized correlation function $c(\Delta \mathbf{r})$ [as in Eq.~\eqref{Eq:Corr_norm} above]. The Born prediction [see Eq.~(2) of the main text] can then be rewritten in the form:

\begin{equation}
\tauSB(\ki, \VR) = \frac{\hbar \ER}{\pi V_R^2}\,\cdot \, \tauSBtilde(\ki) 
\end{equation}
where $\ER=\hbar^2/m\sigma^2$ is the so-called correlation energy and $\tauSBtilde$ is the rescaled scattering time that gives the dependence with the momentum $\ki$. Its expression relies on the integration of the normalized spatial frequency distribution $\tilde{c}(\kdis )=\tilde{c}(\kp -\ki  )$ over the elastic scattering sphere ($|k'|=|k_i|$)~\cite{Rammer2004,Akkermans2007}:
\begin{equation}
\tauSBtilde(\ki) =\frac{4\pi^3 \sigma^2}{|k_i| \, \int d\Omega_{\kp} \, \tilde{c}(\kp -\ki  )},
\end{equation}

\subsection{3D Born prediction $\tauSBtroisD$}

The rescaled Born prediction $\tauSBtroisD$ corresponding to our experimental configuration is shown in Fig.~\ref{fig:Born}. As said above, this calculation uses the effective paraxial model $\CtroisD$ given by Eq.~\eqref{Eq:CorrParaxial} to reproduce the measured two-point correlation function.  
 
 Although the configuration is slightly different, the results discussed in Refs.~\onlinecite{Piraud2012,Piraud2013} for the case of a pure Gaussian illumination still hold for our case~\footnote{In the case of a pure Gaussian illumination, i.e., for $D\rightarrow \infty$, the longitudinal correlation has a pure Lorentzian shape, and the 3D two-point correlation function reads: 
 \begin{equation}
  c_\mathrm{3D,Gauss}=\frac{1}{1+4 \Delta x^2/\sigma_{\|,\mathrm{Gauss} } ^2} \, e^{-\frac{\Delta r_\perp^2/\sigma^2}{1+4 \Delta  x^2/\sigma_{\|,\mathrm{Gauss}}^2} }.
  \label{Eq:Corr_3DGauss}
 \end{equation}
with $\sigma_{\|,\mathrm{Gauss}}=4 \pi \sigma^2/\lambda$.}.
First, $\tauSBtroisD$ increases linearly with the momentum as $\sqrt{\pi} k_i \sigma$ in the large momentum limit ($k_i\gg \sigma^{-1}$), see dash-dotted green line in Fig.~\ref{fig:Born}. This behavior is generic for matter wave and does not depend on the dimension. Second, $\tauSBtroisD$ tends towards a constant in the low momentum limit ($k_i\ll \sigma^{-1}$). This behavior is specific to 
the laser speckle disordered potential~\footnote{In the case of a white noise limit, i.e., finite correlation range, the scattering time increases as $\propto 1/|k_i|$ in the low momentum limit.}: it results from the absence of white noise limit due to the infinite correlation range in the longitudinal direction~\cite{Goodman2007}.

\begin{figure}[!t]
\includegraphics[width=0.48\textwidth]{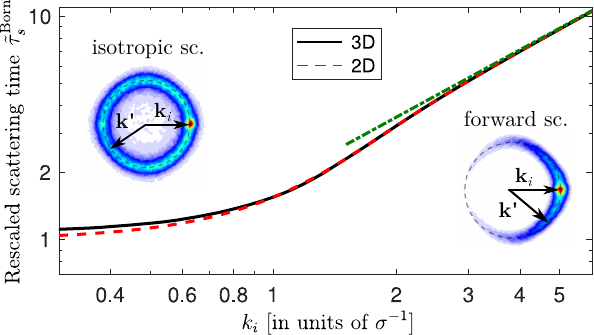}
\caption{\label{fig:Born} \textbf{Rescaled Born prediction $\tauSBtilde$.} Black solid line: 3D Born prediction $\tauSBtroisD$ using the  correlation $\CtroisD(\Delta \mathbf{r})$ given by Eq.~\eqref{Eq:CorrParaxial} (effective paraxial model). Red dashed line: 2D Born prediction $\tauSBdeuxD$ for a transverse Gaussian correlation of size $\sigma$ (1/e radius). Green dash-dotted line: asymptotic behavior at large momentum $\tauSBtilde\sim \sqrt{\pi} k_i \sigma$. Illustrations of the momentum distribution in the isotropic ($k_i\ll \sigma^{-1}$) and forward scattering ($k_i\gg \sigma^{-1}$) regimes are the same as in Fig.~(1) of the main text. }
\end{figure}

\subsection{Comparison with 2D prediction $\tauSBdeuxD$}

The very elongated nature of the laser speckle field, having an infinite correlation range along the longitudinal direction, strongly suggests that our experiment can be described by a pure two-dimensional system. This is confirmed by the excellent agreement between the full 3D calculation $\tauSBtroisD$ and the  Born prediction $\tauSBdeuxD$ for a 2D disordered potential having a Gaussian correlation of size $\sigma$ ($1/e$ radius). In the latter case, one has~\cite{Shapiro2012,Piraud2013}:
\begin{equation}
\tauSBdeuxD(\ki)=e^{  k_i^2 \sigma^2/2 } / I_0 ( k_i^2 \sigma^2/2 )\; , 
\label{Eq:Born2D}
\end{equation}
where $I_0$ is the zero-order modified Bessel function.  $\tauSBdeuxD$ tends towards $1$ for low initial momentum ($k_i\ll \sigma^{-1}$) and, as expected, increases in the same way as the 3D case at large momentum ($\tauSB\sim \sqrt{\pi} k_i \sigma$ for $ k_i\gg \sigma^{-1}$).

\section{Numerical simulations}

The numerical calculations are performed by solving the Schr{\"o}dinger equation for a particle of mass $m$ in a 2D disordered potential $V(\mathbf{r})$. The initial state is a Gaussian wave packet of central momentum $k_i$ and negligible momentum spread $\Delta k$
~\footnote{We have checked that this momentum spread has no influence on the extracted scattering time.}. The scattering time $\tauS$ is extracted from the decay of the initial momentum distribution in the same way than for the experimental data (see Fig.~\ref{fig:tauS_extraction}). The simulations are averaged over 14 different disorder realizations, which is found sufficient to achieve convergence.

\begin{figure}[!b]
\includegraphics[width=0.42\textwidth]{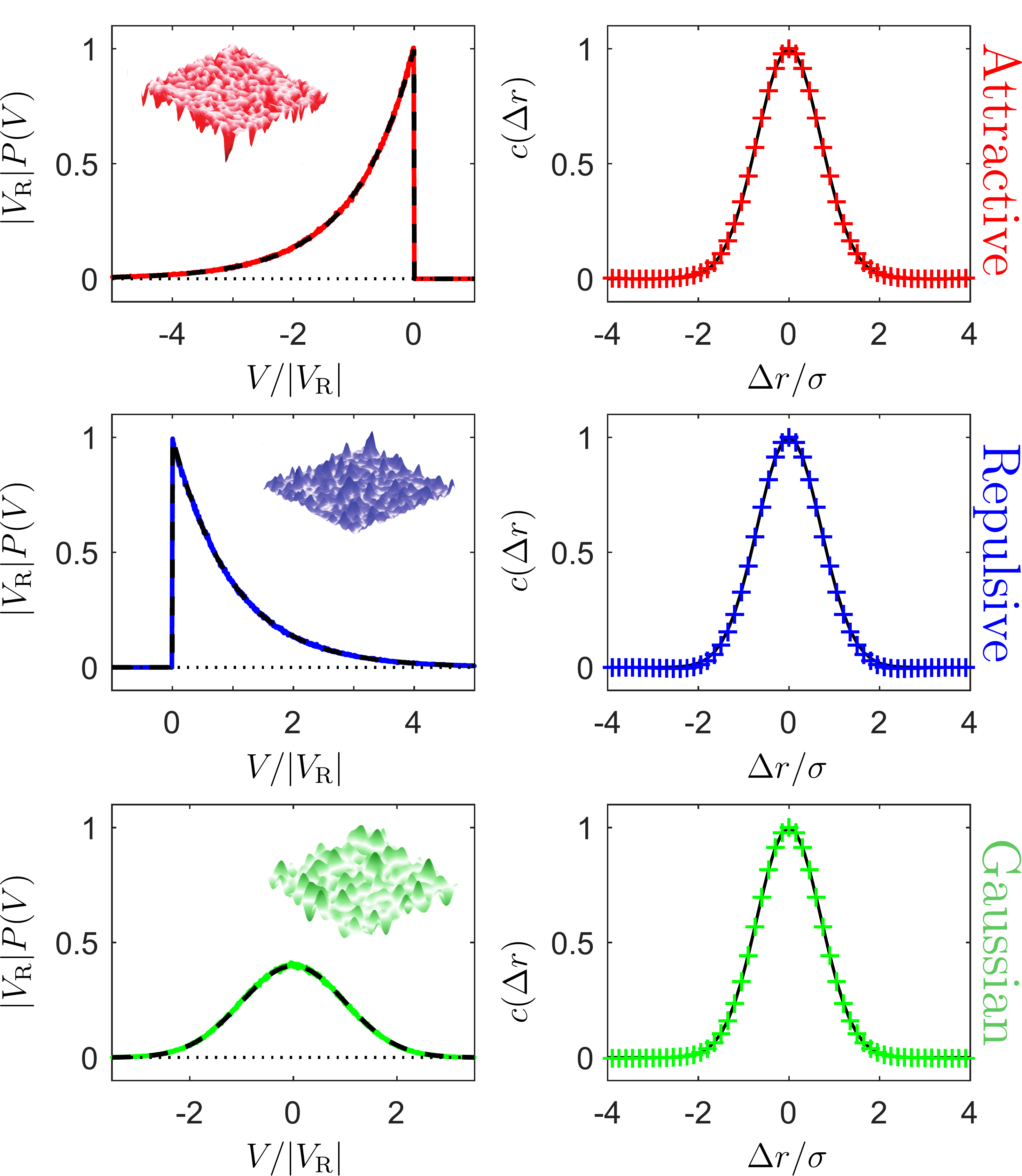}
\caption{\label{fig:Num} \textbf{Numerically generated disorders.} The amplitude probability distribution $P(V)$ (left column) and the normalized correlation function $c(\Delta r)$ (right column) are shown for attractive (first row), repulsive (second row) and Gaussian-distributed (third row) disordered potentials. The amplitude probability distributions are in perfect agreement with the theoretical functions (black dashed lines), namely an exponential function for the attractive and repulsive disorders and a Gaussian function for the Gaussian disorder. The three disorders exhibit the same correlation function, which corresponds to a Gaussian function of size $\sigma$ ($1/e$ radius). }
\end{figure}

To generate the disordered potential, we first calculate the field resulting from the convolution of a spatially uncorrelated complex random field, whose real and imaginary parts are independent Gaussian random variables, with a Gaussian profile accounting for the spatial correlations. The Gaussian-distributed potential is obtained by considering the real part of this field, leading to a Gaussian amplitude probability distribution. Instead, the laser speckle disordered potentials are obtained by considering the intensity (modulus square) of the resulting field~\cite{Goodman2007}, with a negative (resp. positive) sign for the attractive (resp. repulsive) disorder.

In each case, we adjust the amplitude of the complex random field for the rms value of the probability distribution to be $\VR$. It yields $P(V) = |\VR|^{-1} e^{-V / \VR} \cdot \Theta (V/ \VR ) $, with $\Theta$ the step function, for the attractive and repulsive laser speckle fields (first and second rows in Fig.\ref{fig:Num}), and $P(V)=(\sqrt{2\pi}\VR)^{-1} e^{-V^2/2\VR^2}$  for the Gaussian-distributed disorder (3rd row). In the same way, the spatial width of the Gaussian profile is adjusted in each case for the two-point correlation function of the disorder to be a Gaussian of size $\sigma$ ($1/e$ radius), i.e., $c(\Delta r)=e^{-\Delta r^2/\sigma^2}$.


%